\documentclass{aa}

\usepackage{graphicx}
\usepackage{natbib}
\bibpunct{(}{)}{;}{a}{}{,}
\usepackage{times}

\def\arcsec{''}
\def\degr{$^{\circ}$}
\def\Msun{$M_{\odot}$}

\begin{document}
\title{Constraining the difference in convective blueshift between the components of $\alpha$ Cen with precise radial velocities}
\titlerunning{Convective blueshift of $\alpha$ Cen}

\author{D.~Pourbaix\inst{1,11}\fnmsep\thanks{Post-doctoral Researcher, F.N.R.S., Belgium}
\and 
D.~Nidever\inst{2}
\and
C.~McCarthy\inst{3}
\and
R.P.~Butler\inst{3}
\and 
C.G.~Tinney\inst{4}
\and
G.W.~Marcy\inst{5}
\and
H.R.A.~Jones\inst{6}
\and
A.J.~Penny\inst{7}
\and
B.D.~Carter\inst{8}
\and
F.~Bouchy\inst{9}
\and
F.~Pepe\inst{9}
\and
J.B.~Hearnshaw\inst{10}
\and
J.~Skuljan\inst{10}
\and
D.~Ramm\inst{10}
\and
D.~Kent\inst{10}}

\institute{Institut d'Astronomie et d'Astrophysique, Universit\'e Libre de Bruxelles, C.P.~226, Boulevard du Triomphe, B-1050 Bruxelles, Belgium
\and
Department of Physics and Astronomy, San Francisco State University, CA 94132, USA
\and
Department of Terrestrial Magnetism, Carnegie Institution of Washington, 5241 Broad Branch Road NW, Washington DC 20015-1305, USA
\and
Anglo--Australian Observatory, P.O. Box 296, Epping, NSW 1710, Australia
\and
Department of Astronomy, University of California, Berkeley, CA 94720, USA
\and
Astrophysics Research Institute, Liverpool John Moores University, Twelve Quays House, Egerton Wharf Birkenhead CH41 1LD, UK
\and
Rutherford Appleton Laboratory, Chilton, Didcot, Oxon, OX11 0QX, UK
\and
Faculty of Sciences, University of Southern Queensland, Toowoomba, 4350, Australia
\and
Observatoire de Gen\`eve, CH-1290 Sauvergny, Switzerland
\and
Mount John University Observatory, Department of Physics and Astronomy, University of Canterbury, Orivate Bag 4800, Christchurch, New Zealand
\and
Department of Astrophysical Sciences, Princeton University, Princeton NJ 08544-1001, USA
}
\date{Received date; accepted date} 
\offprints{pourbaix@astro.ulb.ac.be}
\abstract{New radial velocities of $\alpha$ Cen A \& B obtained in the framework the Anglo-Australian Planet Search programme as well as in the CORALIE programme are added to those by \citet{Endl-2001:a} to improve the precision of the orbital parameters.  The resulting masses are $1.105\pm0.0070$\Msun\ and $0.934\pm0.0061$\Msun\ for A and B respectively.  The factors limiting how accurately these masses can be derived from a combined visual-spectroscopic solution are investigated.  The total effect of the convective blueshift and the gravitational redshift is also investigated and estimated to differ by $215\pm8$ m/s between the components.  This suggests that the difference in convective blueshift between the components is much smaller than predicted from current hydrodynamical model atmosphere calculations.
}

\maketitle

%
\section{Introduction}
%

\citet{Pourbaix-1999:f} derived an orbit of $\alpha$ Cen A \& B whose main purpose was to illustrate how efficient the combination of visual and spectroscopic data can be in terms of distance and mass determination \citep{Pourbaix-1998:a}.  This system also offers a laboratory for several areas of astronomy ranging from stellar evolution \citep{Pourbaix-1999:b,Guenther-2000:a,Morel-2000:a} and asteroseismology \citep{Bouchy-2001:a} to extra-solar planet search \citep{Murdoch-1993:b,Endl-2001:a}.

Instead of assuming an orbit and looking for a tiny variation caused by an unseen companion in some highly precise radial velocities, we use the data to improve the orbital parameters.  The limits of such an approach are discussed in Sect.~\ref{Sect:Accuracy}.  We improve the spectro-visual orbit in Sect.~\ref{Sect:Orbit} by combining the published data with new ones from the Anglo-Australian \citep{Butler-2001:a} and CORALIE \citep{Queloz-2001:a} planet search projects.  The corrections applied to these data are interpreted in Sect.~\ref{Sect:Shifts} in terms of convective blueshift and gravitational redshift.

%
\section{The parallax as a model parameter}\label{Sect:Accuracy}
%

$\alpha$ Cen is so close to us that there are several sources for the different ingredients required to derive the individual masses, for instance the mass sum and the mass ratio.  We have already argued \citep{Pourbaix-1998:a} that double-lined spectroscopic visual binaries offer a hypothesis-free determination of the distance and individual masses.  The discrepancy between the former orbital parallax $737.0\pm2.6$ mas \citep{Pourbaix-1999:b} and several others $742\pm1.42$ mas \citep[\object{HIP 71683},~][]{Hipparcos}, $747.1\pm1.2$ mas \citep{Soderhjelm-1999} and the consequences on the masses are therefore rather puzzling \citep{Guenther-2000:a}.  

\begin{figure}[htb]
\resizebox{\hsize}{!}{\includegraphics{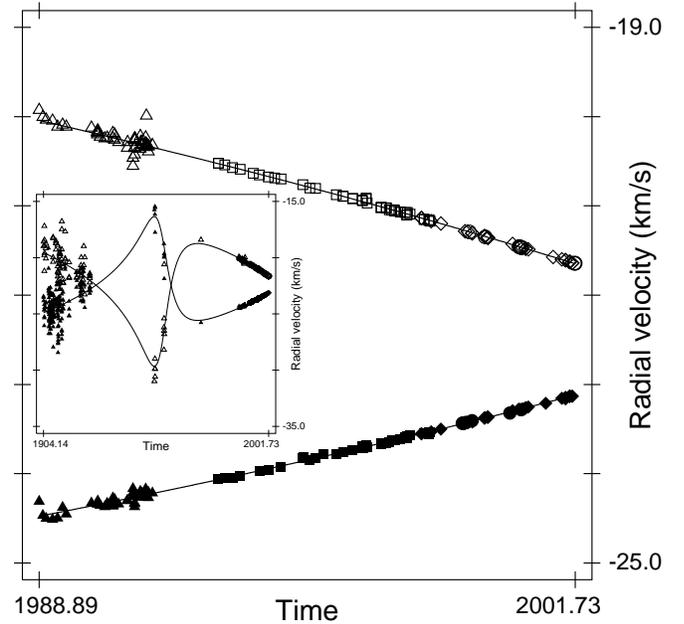}}
\caption[]{\label{Fig:spectplot}Plot of the orbit from Table \ref{Tab:elements} and all the radial velocities obtained over the past 15 years.  The five groups of very precise data are those of \citet{Murdoch-1993:b} (triangles), \citet{Endl-2001:a} (squares), the AAT (diamonds) and CORALIE (circles) ones.  The window shows the orbit with all the existing radial velocities.}
\end{figure}

Even if the old radial velocities exhibit a large scatter (the left end of the window in Fig.~\ref{Fig:spectplot}), in 1999, the precision of $K_A$ and $K_B$ (the amplitude of the radial velocity curves of A and B respectively) was better than 130 m/s for both components thanks to the sole data of \citet{Murdoch-1993:b}.  But how {\em accurate} was the latter set?  \citet{Pourbaix-1999:f} adopted their radial velocity zero point and shifted the earlier sets accordingly.  With the semi-major axis of the visual orbit and the mass-ratio set, parallaxes of 737 mas and 750 mas, i.e. a change of 1.76\%, are distinguishable if $K_A+K_B$ is known to better than 170 m/s.  Unfortunately, the accuracy quoted by \citet{Murdoch-1993:a} was only 200 m/s.  

\begin{figure}[tb]
\resizebox{\hsize}{!}{\includegraphics{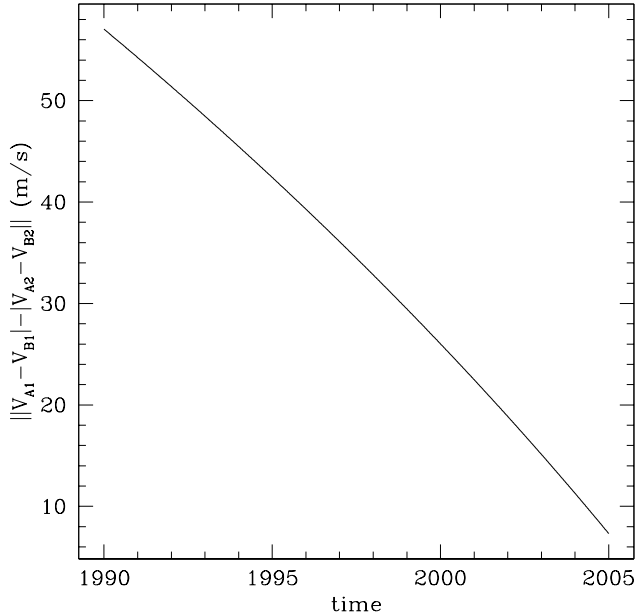}}
\caption[]{\label{Fig:diffrv}Difference of the radial velocity difference between the two components derived from two parallaxes: 747.1 mas \citep{Soderhjelm-1999} and 737 mas \citep{Pourbaix-1999:b}.  $V_{A1}$ \& $V_{B1}$ (resp. $V_{A2}$ \& $V_{B2}$) are the computed radial velocities from the orbit after \citet{Pourbaix-1999:b} adopting 737 mas (resp. 747.1 mas) for the parallax.}
\end{figure}

The difference between $K_A+K_B$ derived with the $747.1\pm1.2$ mas \citep{Soderhjelm-1999} and 737 mas is 169 m/s.  However, the closer one gets to the systemic velocity (Sep. 2006), the smaller the difference in the relative velocity (Fig.~\ref{Fig:diffrv}).  Over the recent observing campaigns (about 5 years long), the effect of this parallax difference was too small to be detected even if accurate absolute radial velocities had been measured.

With the very precise relative radial velocities available now (Sect.~\ref{Sect:Orbit}), one can no longer consider the parallax as an independent parameter of the model.  Indeed, the orbital parallax depends on the corrections applied to these velocities to change them into absolute ones.  Therefore, either one assumes these shifts to be known and derives an orbital parallax or one assumes the latter and derives the former.  From now on, we will assume that the parallax is known and set to $747.1\pm1.2$ mas \citep{Soderhjelm-1999}.
 
%
\section{New data and orbital solution}\label{Sect:Orbit}
%

%
\subsection{AAT data}\label{SubSect:AAT}
%

The relative velocities of A and B were calculated in the usual way \citep{Butler-2001:a}, and have typical internal measurement errors of 3 m/s.  The RMS to a linear fit for A and B velocities is 5.8 and 3.6 m/s.  The absolute velocity zero-point was then derived by reanalyzing the A and B observations using the NSO solar atlas \citep{SoFlAt} as the stellar template \citep{Nidever-2002:a}.

These two sets of radial velocities are absolute in some sense since $V_A$ and $V_B$ have the same zero point. However, neither the AAT nor the CORALIE data are absolute per se since they only barely account for the gravitational redshift and the convective blueshift.  Although these two shifts are rather tough to derive, their total effect can be guessed.

Owing to the similarity between the template and component A, the radial velocities of the latter were assumed to be absolute.  The required correction to the velocities of B was $-215\pm8$ m/s (presumably to account for the gravitational redshift, the convective blueshift and the template mismatch) in order for the orbital parallax to match the adopted value.  This correction has a linear effect on the orbital parallax.  Therefore, the 8 m/s uncertainty corresponds to the 1.2 mas uncertainty of the parallax after \citet{Soderhjelm-1999}.

%
\subsection{CORALIE data}\label{SubSect:CORALIE}
%

The CORALIE data were obtained in the usual way, i.e. a synthetic K0V template was used for both A (slightly evolved G2V) and B (K1V).  The RMS to a linear fit for A and B velocities is 3.15 and 2.54 m/s.  These data together with the visual observations lead to a parallax of 694 mas, about 7\% too small with a mass ratio also discrepant with respect to previous estimates \citep{Kamper-1978,Pourbaix-1999:b}.  By adding corrections of $+9$ m/s and $-215$ m/s to the CORALIE data of A and B respectively, the parallax becomes consistent with the estimate after \citet{Soderhjelm-1999}.  A shift of $+40$ m/s was also needed to put the CORALIE data in the same frame as the AAT one.

We note that for component A the corrections required for the AAT and CORALIE velocities differ by 9 m/s.  This difference of 9 m/s for A could be caused by the use, in the CORALIE analysis, of the K0V template for the G2V star.  The resulting spectral-type mismatch apparently causes a zero-point error of 9 m/s in A.  Actually, although these \mbox{9 m/s} are real, they are likely to be actually spread over the two components, caused by a template mismatch for the two components for both AAT and CORALIE.

Hence, although the instruments, calibration, and reduction techniques are totally different, the AAT and CORALIE velocity sets both suggest that the sum of gravitational redshift and convective blueshifts are different by $\sim$220 m/s between A and B.  We consider this difference of $\sim220$ m/s to be a significant measurement, albeit differential, of the effects of gravitational redshift and convective blueshift in G2V and K1V stars.

%
\subsection{Simultaneous solution}\label{SubSect:orbit}
%

The radial velocities published by \citet{Endl-2001:a} are relative radial velocities.  Therefore, the only information they constrain is the mass ratio (Eq.~\ref{Eq:kappa}).  Once the absolute scale is set, the relative radial velocities can be adjusted accordingly.  The whole set of spectroscopic data is thus composed of the data after \citet{Pourbaix-1999:b}, by \citet{Endl-2001:a} and those in Table \ref{Tab:data}.  A differential correction of $-43$ m/s was also added to the data of B after \citet{Murdoch-1993:b}.  This correction is well within the accuracy limits quoted by these authors.  The zero point of these radial velocities was shifted by $-568$ m/s to coincide with the AAT one.  Consequently, all the older radial velocities were also shifted by $-568$ m/s since they were initially tied \citep{Pourbaix-1999:f} to the data of \citet{Murdoch-1993:b}.  However, the relative weight of these latter data sets is so small with respect to the AAT and CORALIE ones that the effect of these changes is barely noticeable.

The relative position of the secondary has also been measured a few times over the past 15 years.  These observations have been added to the Washington Double Star Catalog of Observations and kindly supplied to us (Hartkopf \& Mason, priv. comm.).  These data combined with all the radial velocities yield the orbit in Table \ref{Tab:elements}.

Although the system exhibits a large proper motion (3.7 \arcsec/yr \citep{Hipparcos}), our model does not account for any perspective effect.  Although the visual observations cover more than three revolutions, they are not precise enough ($\sigma_{\theta}=2.3\deg$ and $\sigma_{\rho}=0.6$\arcsec) to exhibit any noticeable trend.  Even if such a perspective effect should be accounted at the level of precision for the modern radial velocities, the time interval they cover is too short for any noticeable effect.

\begin{table}[htb]
\caption[]{\label{Tab:elements}Orbital parameters and their standard deviations.  The parallax is adopted from \citet{Soderhjelm-1999}}
\begin{tabular}{lcl}\hline
Element                    & Value \\ \hline
$a$ (\arcsec)              & $17.57\pm0.022$\\
$i$ (\degr)                & $79.20\pm0.041$\\
$\omega$ (\degr)           & $231.65\pm0.076$\\
$\Omega$ (\degr)           & $204.85\pm0.084$\\
$e$                        & $0.5179\pm0.00076$\\
$P$ (yr)                   & $79.91\pm0.011$\\
$T$ (Besselian year)       & $1875.66\pm0.012$\\
$V_0$ (km/s)               & $-22.445\pm0.0021$\\
$\varpi$ (mas)             & $747.1\pm1.2$ & (adopted)\\
\medskip
$\kappa=M_B/(M_A+M_B)$     & $0.4581\pm0.00098$\\

$M_A$ (\Msun)              & $1.105\pm0.0070$\\
$M_B$ (\Msun)              & $0.934\pm0.0061$\\
\hline
\end{tabular}
\end{table}

\begin{table*}[htb]
\caption[]{\label{Tab:data}Relative and semi-absolute radial velocities}
\begin{tabular}{lrrrrrrrrl}\hline
JD & $\Delta_A$ & $\sigma_{\Delta_A}$ & $\Delta_B$ & $\sigma_{\Delta_B}$ & $\stackrel{\circ}{V_A}$ &$\sigma_{V_A}$ & $\stackrel{\circ}{V_B}$ & $\sigma_{V_B}$ & Reference\\   
-2,400,000 & (m/s) & (m/s) & (m/s) & (m/s) & (m/s) & (m/s) & (m/s) & (m/s) & \\ \hline   
50831.27 & -339.6 & 2.1 & 224.7 & 1.7 & -23552.3 & 2.1 & -20920.2 & 1.7 & AAT\\
50833.24 & -338.0 & 2.4 & 223.0 & 2.0 & -23550.7 & 2.4 & -20921.9 & 2.0 & AAT\\
50917.12 & -317.5 & 3.9 & 188.0 & 4.4 & -23530.2 & 3.9 & -20956.9 & 4.4 & AAT\\
51002.93 & -287.8 & 3.0 & 163.8 & 2.9 & -23500.5 & 3.0 & -20981.1 & 2.9 & AAT\\
51189.88 & & & & & -23487.1 & $\le2$ & & &CORALIE\\
51193.89 & & & & & -23483.5 & $\le2$ & & &CORALIE\\
51213.28 & -228.2 & 2.6 & 77.3 & 3.2 & -23440.9 & 2.6 & -21067.6 & 3.2 & AAT\\
51223.87 & & & & & -23474.2 & $\le2$ & & &CORALIE\\
51236.29 & -214.7 & 3.7 & 73.4 & 3.0 & -23427.4 & 3.7 & -21071.5 & 3.0 & AAT\\
51251.86 & & & & & & & -21115.0 & $\le2$ &CORALIE\\
51274.23 &        &     & 54.5 & 3.3 &          &     & -21090.5 & 3.3 & AAT\\
51274.75 & & & & & -23455.7 & $\le2$ & & &CORALIE\\
51276.08 & -209.3 & 3.3 & 54.7 & 3.2 & -23422.0 & 3.3 & -21090.2 & 3.2 & AAT\\
51382.86 & -163.5 & 3.1 & 21.0 & 2.8 & -23376.2 & 3.1 & -21123.9 & 2.8 & AAT\\
51383.86 & -162.4 & 3.0 & 18.6 & 3.0 & -23375.1 & 3.0 & -21126.3 & 3.0 & AAT\\
51385.85 & -162.1 & 3.3 & 14.7 & 3.1 & -23374.8 & 3.3 & -21130.2 & 3.1 & AAT\\
51386.84 & -159.0 & 2.9 & 20.8 & 3.0 & -23371.7 & 2.9 & -21124.1 & 3.0 & AAT\\
51387.47 & & & & & & & -21174.4 & $\le2$ &CORALIE\\
51410.86 & -153.1 & 2.3 & 7.4 & 3.9 & -23365.8 & 2.3 & -21137.5 & 3.9 & AAT\\
51412.85 & -156.8 & 3.8 & 5.0 & 3.0 & -23369.5 & 3.8 & -21139.9 & 3.0 & AAT\\
51413.85 & -152.1 & 4.3 & 0.0 & 4.3 & -23364.8 & 4.3 & -21144.9 & 4.3 & AAT\\
51606.87 & & & & & -23368.3 & $\le2$ & & &CORALIE\\
51630.31 & -81.5 & 2.1 & -67.5 & 3.2 & -23294.2 & 2.1 & -21212.4 & 3.2 & AAT\\
51683.01 & -67.2 & 3.7 & -93.7 & 3.2 & -23279.9 & 3.7 & -21238.6 & 3.2 & AAT\\
51684.09 & -64.9 & 3.7 & -99.4 & 2.9 & -23277.6 & 3.7 & -21244.3 & 2.9 & AAT\\
51686.57 & & & & & & & -21281.3 & $\le2$ &CORALIE\\
51687.74 & & & & & & & -21282.1 & $\le2$ &CORALIE\\
51688.67 & & & & & & & -21285.4 & $\le2$ &CORALIE\\
51698.46 & & & & & & & -21289.7 & $\le2$ &CORALIE\\
51703.47 & & & & & & & -21292.6 & $\le2$ &CORALIE\\
51704.47 & & & & & -23331.7 & $\le2$ & & &CORALIE\\
51704.48 & & & & & & & -21292.3 & $\le2$ &CORALIE\\
51707.47 & & & & & & & -21287.2 & $\le2$ &CORALIE\\
51717.83 & -63.4 & 4.2 & -111.0 & 3.7 & -23276.1 & 4.2 & -21255.9 & 3.7 & AAT\\
51742.88 & -49.7 & 2.2 & -117.9 & 2.8 & -23262.3 & 2.2 & -21262.8 & 2.8 & AAT\\
51743.83 & -46.3 & 3.2 & -117.7 & 3.0 & -23259.0 & 3.2 & -21262.6 & 3.0 & AAT\\
51766.87 & -39.8 & 2.2 & -127.1 & 2.7 & -23252.5 & 2.2 & -21272.0 & 2.7 & AAT\\
51919.28 &   2.0 & 0.6 &        &     & -23210.7 & 0.6 &  &  & AAT\\
51984.32 &       &     & -211.4 & 1.6 &          &     & -21356.3 & 1.6 & AAT\\
52061.01 & 58.7 & 3.8 & -239.2 & 3.6 & -23154.0 & 3.8 & -21384.1 & 3.6 & AAT\\
52091.93 & 51.5 & 4.0 & -249.7 & 3.5 & -23161.2 & 4.0 & -21394.6 & 3.5 & AAT\\
52093.00 & 69.8 & 3.7 & -247.8 & 3.3 & -23142.9 & 3.7 & -21392.7 & 3.3 & AAT\\
52126.89 & 79.0 & 1.8 & -269.4 & 2.8 & -23133.7 & 1.8 & -21414.4 & 2.8 & AAT\\
52127.91 & 77.8 & 2.7 & -265.4 & 3.1 & -23134.9 & 2.7 & -21410.3 & 3.1 & AAT\\
52129.02 & 79.8 & 3.9 & -270.2 & 3.4 & -23132.9 & 3.9 & -21415.1 & 3.4 & AAT\\
52129.96 & 82.1 & 3.4 & -265.9 & 3.1 & -23130.6 & 3.4 & -21410.8 & 3.1 & AAT\\
52150.94 & 81.6 & 2.8 & -280.0 & 2.8 & -23131.1 & 2.8 & -21425.0 & 2.8 & AAT\\
52151.85 & 81.9 & 2.5 & -278.6 & 2.0 & -23130.8 & 2.5 & -21423.5 & 2.0 & AAT\\
52171.49 & & & & & & & -21467.7 & $\le2$ &CORALIE\\
52173.50 & & & & & & & -21466.6 & $\le2$ &CORALIE\\
52177.49 & & & & & & & -21468.0 & $\le2$ &CORALIE\\
\hline
\end{tabular}
\end{table*}

With respect to the 1999 solution, one notes an upward revision of the fractional mass (still rather consistent with the estimate after \citet{Murdoch-1993:a}) which is hypothesis-free and relies on the relative velocities only (Sect.~\ref{Sect:Shifts}).  This change combined with the larger parallax yield smaller masses, with a larger effect on A than on B.  The precision on the masses ($<1$\%), about 4.5 times smaller than in 1999, therefore reflects the precision of the parallax and the accuracy of the mass ratio, orbital period and semi-major axis of the visual orbit.  The revised mass of the primary agrees more with what was assumed in some recent theoretical investigations \citep[e.g.][]{Guenther-2000:a}.  It is also quite consistent with the recent result of asteroseismology by \citet{Carrier-2001:a}.

With respect to the orbit, the radial velocities of \citet{Endl-2001:a} have standard deviations of 11.9 m/s and 12.4 m/s for A and B respectively, quite consistent with the mean measurement errors (12.3 m/s and 9.9 m/s).  For the AAT data, the standard deviations are 4.8 m/s and 3.7 m/s, again very consistent with the estimated mean measurement errors.  The CORALIE radial velocities have standard deviations of 7.6 and 4.3 m/s for A and B respectively.  For this latter set, the larger noise on component A is due to some guiding problems noticed for bright stars.

%
\section{Blue and red shifts}\label{Sect:Shifts}
%

Radial velocities of double lined spectroscopic binaries are often used to derive the mass ratio \citep{Irwin-1973}.  As long as the precision of these velocities does not exceed their accuracy, this is indeed a licit approach.  Nowadays, the radial velocities have a precision better than 10 m/s, however there has been little concern since \citet{Murdoch-1993:b} about improving their accuracy.  Indeed, whereas the precision is essential for the planet search campaign, very few people seem to worry about the accuracy \citep{GriffinREM-1999:a,Gullberg-1999:a}.

Starting with the equations of the radial velocity curves \citep{PrDoSt},
\begin{equation}
\begin{array}{l}
V_A=V_0-K_A(\cos(\omega+v)+e\cos\omega),\\
V_B=V_0+K_B(\cos(\omega+v)+e\cos\omega)
\end{array}
\end{equation}
where $v$ is the true anomaly and $\omega$ is the argument of the periastron of B, one usually derives
\begin{equation}\label{Eq:kappa}
{V_B-V_0 \over V_0-V_A}={K_B\over K_A}={M_A\over M_B}={1\over\kappa}-1
\end{equation}
where $V_A$ and $V_B$ represent the absolute radial velocities of both components, not the observed ones.  Indeed, the latter are affected by the gravitational redshift \citep{AsCo} and the convective blueshift \citep{Dravins-1999:a}.  Therefore, unless the measured radial velocities are corrected for the total effect of these two shifts, these velocities do not yield the mass ratio.

If the mass and radius of the star are known, the gravitational redshift can be computed \citep{AsCo}.  Even at a 10 m/s precision, the differential effect of that shift would already be rather large.  Thus, with the stellar parameters adopted below, the differential redshift is 143 m/s, i.e.\ a $14\sigma$ effect.  Accounting for the convective blueshift is much more difficult because that would require some {\em a priori} understanding of the internal structure of the star.

Actually, one does not need (and should even avoid) the absolute radial velocities to derive the mass ratio.  Indeed, the latter can already be obtained from the relative radial velocities.  From
\begin{equation}
\begin{array}{l}
V_A=V_0+V_{0A}+\Delta_A\\
V_B=V_0+V_{0B}+\Delta_B
\end{array}
\end{equation}
where $V_{0A}$ and $V_{0B}$ account for the zero points and the gravitational and convective shifts and $\Delta_A$ and $\Delta_B$ are the relative radial velocities, one derives
\begin{equation}
\kappa={M_B\over M_A+M_B}={\Delta_A\over\Delta_A-\Delta_B}.
\end{equation}
Thus, from \citet{Endl-2001:a}, an orthogonal least-square fit yields $\kappa=0.457$ which is very consistent with $0.454\pm0.002$ after \citet{Kamper-1978}.

We showed in the previous section that both the CORALIE and AAT data had to be differentially corrected by about \mbox{220 m/s} in order to obtain likely parallax and mass ratio.  We can therefore write
\begin{equation}
(G_B-C_B)-(G_A-C_A)=215\pm8 {\rm ~m/s}
\end{equation}
where $G$ denotes the gravitational redshift and $C$ the convective blueshift (both regarded as positive quantities).  It is worth keeping in mind that the 8 m/s uncertainty is based on the AAT data, results from the uncertainty on the adopted parallax only and does not account for the possible effect of the template mismatch.  With the effective temperatures from \citet{Neuforge-1997}, the luminosities from \citet{Guenther-2000:a} and the masses we obtain (Table \ref{Tab:elements}), $G_A=560\pm14$ m/s and $G_B=703\pm20$ m/s thus yielding
\begin{equation}
C_A-C_B=72\pm26 {\rm ~m/s}
\end{equation}
for the overall spectrum.  The difference in convective blue\-shifts between the components of this system for the Fe I line at $\lambda 520$ nm is about 500 m/s \citep{Dravins-1990:a}.  Hence, for at least one of these two stars, the Fe I line does not seems to be a good indicator for the overall convective blueshift.

%
\section{Conclusion}
%

We have solved the 3D two-body problem to derive masses and their precisions.  The advantage of this solution is that it is completely independent of theoretical considerations such as models of stellar structure and evolution.  Therefore, unless the original observations (the relative positions or the radial velocities) become questionable, these masses should be considered as true observables and the parameters of the theoretical models adjusted accordingly (not the opposite).  However, in the case of double-lined spectroscopic visual binaries, although the mass ratio is accurately derived, only the precision can be guaranteed for the individual masses.

We have studied in detail the case of $\alpha$ Cen A \& B.  Thanks to the accurate estimate of the parallax (e.g. after Hipparcos) and the combination of precise radial velocities and positions, it is possible to constrain the total effect of the convective blueshift and the gravitational redshift.

\begin{acknowledgements}
We thank L.~Lindegren, the referee, for his valuable suggestions.
The Anglo-Australian Planet Search team would like to thank the
Director of the AAO, B.~Boyle, and the superb technical support
of the AAT staff -- in particular E.~Penny, R.~Patterson, D.~Stafford,
F.~Freeman, S.~Lee, J.~Pogson and G.~Schafer.  We acknowledge the UK
and Australian government support of the AAT through their PPARC and
DETYA (HRAJ, CGT, AJP) ; NASA grant NAG5-8299, \& NSF grant AST95-20443
(GWM); NASA grant NAG5-11094 (DP); NSF grant AST-9988987 (RPB) and Sun 
Microsystems.
\end{acknowledgements}

%
\bibliographystyle{apj}
\bibliography{articles,books}
%

\end{document}